\documentclass[aps,showpacs,twocolumn,longbibliography,nofootinbib,secnumarabic,superscriptaddress]{revtex4-1}
\usepackage{graphicx}
\usepackage{array}
\usepackage{color}
\usepackage{amsmath}
\usepackage{amsxtra}
\usepackage{amstext}
\usepackage{amssymb}
\usepackage{latexsym}
\usepackage{verbatim}
\usepackage{comment}
\usepackage{epstopdf}

\begin{document}

\title
%{Single-mode nonclassicality \\ as Holstein-Primakoff transformation of two-mode entangled ${\cal N}$ identical particles  }
{Measuring nonclasicality of single-mode systems}
%\author{Deniz T\"{u}rkpen\c{c}e}
%\affiliation{Department of Electrical \& Electronics Engineering, K{\i}rklareli University, 39100 Kayal{\i}, K{\i}rklareli, Turkey}
%%\affiliation{Center for Advanced Researches, K{\i}rklareli University, 39020 Karah{\i}d{\i}r, K{\i}rklareli, Turkey}
%\author{G\"{u}rsoy B. Akg\"{u}\c{c}}
%\affiliation{Department of Physics, Bilkent University, 06800, Ankara, Turkey}
%\author{Alpan Bek}
%\affiliation{Department of Physics, Middle East Technical University, 06800, Ankara, Turkey}
\author{Mehmet Emre Tasgin}
\affiliation{Institute of Nuclear Sciences, Hacettepe University, 06800, Ankara, Turkey}
%\affiliation{to whom correspondence should be addressed.}

\date{\today}

\begin{abstract}
A beam splitter can generate entanglement if the input field is nonclassical, e.g. squeezed. Thus, beam splitter transformation can be utilized for obtaining single-mode nonclassicality conditions from two-mode entanglement criteria. Here, we demonstrate that (i) the generalized quadrature-squeezing condition $\langle\hat{a}^\dagger\hat{a}\rangle < |\langle \hat{a}^2\rangle|$ can be obtained from the two-mode entanglement criterion of Duan-Giedke-Cirac-Zoller~[PRL 84, 2722]. (ii) Single-mode nonclassicality condition, obtained from the two-mode entanglement criterion of Simon~[PRL, 84 2726], can detect (also accompanies) the nonclassicality of a superradiant ground state where no quadrature-squeezing takes place. (iii) We present an explicit form for a single-mode nonclassicality measure, for Gaussian states, via maximization of the beam splitter output with respect to the beam splitter parameters. We measure the nonclassicality of a squeezed state and the one generated by a damped second harmonic crystal cavity. We determine pump strengths where single-mode nonclassicality onsets, i.e. resists against the vacuum noise.
\end{abstract}

%\pacs{03.67.Bg, 03.67.Mn, 42.50.Dv, 42.50.Ex}

%03.67.Bg : Entanglement and quantum nonlocality in quantum information
%03.67.Mn : Entanglement and quantum nonlocality in quantum information
%42.50.Dv : Photons nonclassical states
%42.50.Ex : Quantum information, optical implementations, 

\maketitle

%%%%%%%%%%%%%%%%%%%%%%%%%%%%%%%%%%%%%%%%%%%%%%%%%%%%%%%%%%%%%%%%%%%%%%%%%%%%%%%%%%%%%%%%%%%%%%%%%%%%%%%%%%%
%%%%%%%%%%%%%%%%%%%%%%%%%%%%%%%%%%%%%%%%%%%%%%%%%%%%%%%%%%%%%%%%%%%%%%%%%%%%%%%%%%%%%%%%%%%%%%%%%%%%%%%%%%%
%%%%%%%%%%%%%%%%%%%%%%%%%%%%%%%%%%%%%%%%%%%%%%%%%%%%%%%%%%%%%%%%%%%%%%%%%%%%%%%%%%%%%%%%%%%%%%%%%%%%%%%%%%%
%%%%%%%%%%%%%%%%%%%%%%%%%%%%%%%%%%%%%%%%%%%%%%%%%%%%%%%%%%%%%%%%%%%%%%%%%%%%%%%%%%%%%%%%%%%%%%%%%%%%%%%%%%%
%%%%%%%%%%%%%%%%%%%%%%%%%%%%%%%%%%%%%%%%%%%%%%%%%%%%%%%%%%%%%%%%%%%%%%%%%%%%%%%%%%%%%%%%%%%%%%%%%%%%%%%%%%%
%%%%%%%%%%%%%%%%%%%%%%%%%%%%%%%%%%%%%%%%%%%%%%%%%%%%%%%%%%%%%%%%%%%%%%%%%%%%%%%%%%%%%%%%%%%%%%%%%%%%%%%%%%%
\section{Introduction}

Achieving nonclassical states of a single-mode light is attractive for realization of genuine quantum optical phenomena. A single-mode nonclassical field can generate entangled photon pairs at the output of a beam-splitter~\cite{Kim&Knight2002,BSentanglement,aharanov1966}, vital for quantum information and quantum teleportation~\cite{teleport1,teleport2,teleport3}. Additionally, single-mode nonclassical fields, e.g. squeezed states, can be utilized to perform measurements below the standard quantum limit~\cite{measurementSQL1,measurementSQL2,measurementSQL3,measurementSQL4} which, for instance, enabled the detection of  gravitational waves~\cite{LIGO2013}.

In the detection of the single-mode nonclassicality, usually, the negativity of quasiprobability distributions, such as Wigner function~\cite{Wnegativity1,Wnegativity2} or Sudarshan-Glauber P distribution~\cite{Pnegativity,SudarshanP,WignerfnxExperiment,measurementSQL3} are evaluated. Detection of single-mode nonclassicality~(SMNc) can be performed by probing negative values of quasi-probability distributions, e.g., Wigner function~\cite{Wnegativity1,Wnegativity2}, where a negative value~\cite{WignerfnxExperiment,measurementSQL3} implies that the state is nonclassical. The well-known detection parameters and witnesses for single-mode nonclassicality, e.g. squeezing, also, rely on the negativity of similar quasiprobability distributions ~\cite{Wnegativity1,Wnegativity2,Pnegativity,SudarshanP}. More explicitly, for instance, if there is quadrature-squeezing $(\Delta x)^2<1/2$, the quasiprobability distribution $P(\alpha)$, in ${\cal N}=(\Delta x)^2-1/2=\int d^2 \alpha P(\alpha) (\alpha-\langle \hat{x}\rangle)^2$, has to be negative at some values of $\alpha$, since $(\alpha-\langle \hat{x}\rangle)^2$ is positive-definite. That is, the single-mode state cannot be composed of classical (coherent) states with a classical probability distribution $P(\alpha)$.

Recently, we demonstrated a new method for obtaining single-mode nonclassicality criteria from many-particle entanglement criteria~\cite{tasgin2017many,tasgin2015HP} via a Holstein-Primakoff~(HP) transformation~\cite{Emary&BrandesPRE2003}. We utilize the simple relation: atomic coherent states~(ACSs), separable many-particle states, mimic the coherent states of light~\cite{
Klauder1985,Radcliffe1971} in the large particle number, $N\to\infty$, limit. In Ref.~\cite{tasgin2015HP}, it is further shown that a two-mode entanglement criterion can also be utilized as a SMNc criterion by using the Holstein-Primakoff transformation. Here the two-mode entanglement is referred to the entanglement between the two levels, $\hat{c}_g$ and $\hat{c}_e$, of an ensemble of identical particles, where $\hat{c}_{g,e}$ annihilates a particle in the ground/excited state.

These latter two approaches, however, have a very limited use. Because, for instance, two major criteria, Simon-Peres-Horodecki~(SPH)~\cite{SimonPRL2000,AdessoPRA2004,WernerPRA2002} and Duan-Giedke-Cirac-Zoller~(DGCZ)~\cite{Duan&Zoller2000}, include terms which do not preserve the number of particles, e.g. $\hat{c}_g\hat{c}_e$. This limits the use of Holstein-Primakoff transformation seriously.

In the past decade, Asboth {\it et al.} \cite{EntanglementPot} brought a new perspective on the concept of nonclassicality. They associated the degree of the nonclassicality of a single-mode field with its ability to produce two-mode entanglement \cite{SimonPRA2011} at the output of a beam-splitter. More recently, Vogel and Sperling \cite{Vogel&Sperling2014,Mraz&Hage2014} demonstrated that rank of two-mode entanglement, a single-mode light generates through a beam-splitter, is equal to the rank of the expansion of this nonclassical state in terms of classical coherent states. Namely, $|\psi_{\rm \scriptscriptstyle Ncl}\rangle=\sum_{i=1}^{r}\kappa_i |\alpha_i\rangle$ generates the state 
$|\psi_{\rm \scriptscriptstyle Ent}\rangle=\sum_{i=1}^{r}\lambda_i |a_1^{(i)}\rangle \otimes |a_2^{(i)}\rangle$,  where $|a_1^{(i)}\rangle$ and $|a_2^{(i)}\rangle$ represent the states of output modes, $\hat{a}_1$ and $\hat{a}_2$, of the beam-splitter.

Hillery and Zubairy established connections between single-mode nonclassicality conditions and two-mode entanglement criteria~\cite{HZPRA2006} utilizing the beam splitter transformation.

In this paper, we present further utilizations of the beam-splitter transformation for achieving single-mode nonclassicality measure/conditions from two-mode entanglement measure/criteria. We obtain a single-mode nonclassicality measure based on the strength of the entanglement at the output of a beam-splitter, i.e. its entanglement potential~\cite{EntanglementPot}. We consider a single-mode state and obtain an explicit form for the strength of the entanglement it generates at the beam-splitter output in terms of $\langle\hat{a}^\dagger\hat{a}\rangle$ and $\langle \hat{a}^2\rangle$, with $\hat{a}$ is the annihilation operator of the single-mode. At the output, we use the logarithmic-negativity~(log-neg, $E_{\cal N}$)~\cite{AdessoPRA2004,WernerPRA2002} to measure the generated entanglement for Gaussian states, where $E_{\cal N}$ is a {\it measure} for Gaussian states~\cite{Plenio2005}. We use $E_{\cal N}$ to measure the nonclassicality of a quadrature-squeezed state. We also measure the nonclassicality created in a second harmonic crystal cavity with coupling to reservoir noise and we determine the pump strengths where single-mode nonclassicality becomes resistant (nonzero) against the environment noise, see Fig.~\ref{fig2}.

For non-Gaussian states, we use the criterion of Simon-Peres-Horodecki~(SPH)~\cite{SimonPRL2000} ---not an entanglement measure. We demonstrate that the criterion, used at the output of a beam-splitter, can detect the single-mode nonclassicality also in the ground state of a superradiant phase transition where the state is non-Gaussian. Furthermore, SPH witness accompanies the phase of the transition, see Fig.~\ref{fig4}.

We calculate the 4$\times$4 noise matrix of the two-mode output state in terms of $\langle\hat{a}^\dagger\hat{a}\rangle$ and $\langle \hat{a}^2\rangle$ and evaluate the log-neg $E_{\cal N}$. We ``maximize" $E_{\cal N}$ with respect to the beam splitter parameters $r$,$t$ and $\phi$, reflection, transmission and phase of the beam splitter. We call this optimum value as ${\cal N}_{\rm \scriptscriptstyle SMNc}$, see Sec.~\ref{sec:SMNc} for details. We also optimize the Simon's criterion which we utilize as a single-mode nonclassicality~(SMNc) condition, i.e. $\eta_{\rm \scriptscriptstyle SMNc}<0$.

We further demonstrate that a single-mode nonclassicality condition, i.e. $\langle\hat{a}^\dagger\hat{a}\rangle< |\langle \hat{a}^2\rangle|$, can be obtained from the DGCZ two-mode entanglement criterion, via an optimization in the parameters of the DGCZ criterion and the the beam splitter. We discuss that this condition, actually, is the generalized quadrature-squeezing condition, which determines the maximum squeezing (minimum noise) in the single-mode system.

The paper is organized as follows. In Sec.~\ref{sec:SPH}, we introduce the two-mode entanglement criterion Simon derived in Ref.~\cite{SimonPRL2000}, SPH, and the logarithmic-negativity~\cite{AdessoPRA2004,WernerPRA2002} which is an entanglement measure for Gaussian states~\cite{Plenio2005}. In Sec.~\ref{sec:SMNc}, we introduce the beam splitter transformation. We obtain the 4$\times$4 two-mode noise matrix, a given single-mode generates, at the beam-splitter output. Thus, we obtain a single-mode nonclassicality measure via $E_{\cal N}$ for Gaussian states and a single-mode condition via using the SPH criterion for non-Gaussian states. In Sec.~\ref{sec:examples}, we test the SMNc measure for a (\ref{sec:squeezing}) quadrature-squeezed state and (\ref{sec:SHG}) on the nonclassicality a second harmonic crystal cavity generates. Then, we show (\ref{sec:Superradiance}) that the SMNC condition ---obtained via SPH citerion--- accompanies the single-mode nonclassicality in a superradiant phase transition. In Sec.~\ref{sec:SMNcfromDGCZ}, we demonstrate that generalized quadrature-squeezing condition $\langle\hat{a}^\dagger\hat{a}\rangle < |\langle \hat{a}^2\rangle|$ can be obtained from DGCZ two-mode entanglement criterion via the beam-splitter approach. In Sec.~\ref{sec:summary}, we present a summary of our results.

To our knowledge, even though some works~\cite{HZPRA2006,MGAParisPRA2015,NhaSciRep2019} have already been carried out on obtaining single-mode nonclassicality conditions from two-mode entanglement criteria, the following important treatments/demonstrations are absent in the literature. Obtaining the matrix forms for a (i,ii) measure/condition in terms of $\langle \hat{a}^2\rangle$ and $\langle\hat{a}^\dagger \hat{a} \rangle$ and carrying out an optimization for (i,ii) single-mode nonclassicality measure/condition, measurement of the nonclasicality of, e.g., second harmonic crystal cavity with dampings, demonstration of such a condition for superradiant states, and (iii) obtaining the generalized quadrature-squeezing condition via the optimization of DGCZ criterion have not been carried out.

%%%%%%%%%%%%%%%%%%%%%%%%%%%%%%%%%%%%%%%%%%%%%%%%%%%%%%%%%%%%%%%%%%%%%%%%%%%%%%%%%%%%%%%%%%%%%%%%%%%%%%%%%%%
%%%%%%%%%%%%%%%%%%%%%%%%%%%%%%%%%%%%%%%%%%%%%%%%%%%%%%%%%%%%%%%%%%%%%%%%%%%%%%%%%%%%%%%%%%%%%%%%%%%%%%%%%%%
%%%%%%%%%%%%%%%%%%%%%%%%%%%%%%%%%%%%%%%%%%%%%%%%%%%%%%%%%%%%%%%%%%%%%%%%%%%%%%%%%%%%%%%%%%%%%%%%%%%%%%%%%%%
\section{Nonclassicality measure from SPH criterion} 

In this section, we introduce the Simon-Peres-Horodecki~(SPH) two-mode entanglement criterion~\cite{SimonPRL2000} and the log-neg entanglement measure $E_{\cal N}$~\cite{AdessoPRA2004,WernerPRA2002}. Next, we relate the variance matrix of the SPH criterion to single-mode noises $\langle \hat{a}^2\rangle$ and $\langle \hat{a}^\dagger\hat{a}\rangle$ using the beam-splitter transformations \cite{Kim&Knight2002,BSentanglement,aharanov1966}. We obtain a measure for quantifying the degree of nonclassicality $E_{\cal N}\big(\langle \hat{a}^2\rangle, \langle \hat{a}^\dagger\hat{a}\rangle\big)$ of Gaussian states. We also obtain a single-mode nonclassicality condition $\lambda_{\scriptscriptstyle \rm Simon}\big(\langle \hat{a}^2\rangle, \langle \hat{a}^\dagger\hat{a}\rangle\big)$ from the Simon's (SPH) two-mode entanglement criterion~\cite{SimonPRL2000}. In the next section, we present calculations for three different systems.

%%%%%%%%%%%%%%%%%%%%%%%%%%%%%%%%%%%%%%%%%%%%%%%%%%%%%%%%%%%%%%%%%%%%%%%%%%%%%%%%%%%%%%%%%%%%%%%%%%%%%%%%%%%
\subsection{Two-mode entanglement} \label{sec:SPH}

A two-mode Gaussian state can be completely characterized by its covariance (correlation) matrix \cite{AdessoPRA2004,WernerPRA2002,optomechEnt1,Simon1988} 
\begin{equation}
V_{ij}=\frac{1}{2} \langle\hat{Y}_i\hat{Y}_j + \hat{Y}_j\hat{Y}_i \rangle - \langle\hat{Y}_i \rangle \langle\hat{Y}_j \rangle  \; ,
\label{Vmatrix}
\end{equation}
where $\hat{Y}=[\hat{x}_1\: , \: \hat{p}_1 \: , \: \hat{x}_2\: , \: \hat{p}_2]$, $\hat{x}_i=(\hat{a}_i^\dagger+\hat{a}_i)/\sqrt{2}$ and $\hat{p}_i={\rm i}(\hat{a}_i^\dagger-\hat{a}_i)/\sqrt{2}$. We note that the variances $\langle\hat{Y}_{i,j} \rangle$ in Eq.~(\ref{Vmatrix}) can be adjusted to zero without affecting the entanglement features \cite{simonPRA1994,AdessoPRA2004}. Correlation matrix can be written as 
\begin{equation}
 V=\left[ 
 \begin{array}{cc}
A   & C  \\
C^T & B \\
\end{array} \right],
\label{VmatrixABC}
\end{equation}
where $A$, $B$, and $C$ are the 2$\times$2 matrices
\begin{equation}
A=\resizebox{.9\hsize}{!}{$
\left[  \begin{array}{cc}
\langle\hat{x}_1^2\rangle - \langle\hat{x}_1\rangle^2  & \langle \hat{x}_1\hat{p}_1 + \hat{p}_1\hat{x}_1\rangle/2 - \langle\hat{x}_1\rangle\langle\hat{p}_1\rangle \\
\langle \hat{x}_1\hat{p}_1 + \hat{p}_1\hat{x}_1\rangle/2 -\langle\hat{x}_1\rangle\langle\hat{p}_1\rangle & \langle\hat{p}_1^2\rangle - \langle\hat{p}_1\rangle^2\\
\end{array} \right] \; , $}
\label{Amat}
\end{equation}
\begin{equation}
B=\resizebox{.9\hsize}{!}{$
\left[  \begin{array}{cc}
\langle\hat{x}_2^2\rangle - \langle\hat{x}_2\rangle^2  & \langle \hat{x}_2\hat{p}_2 + \hat{p}_2\hat{x}_2\rangle/2 - \langle\hat{x}_2\rangle\langle\hat{p}_2\rangle \\
\langle \hat{x}_2\hat{p}_2 + \hat{p}_2\hat{x}_2\rangle/2 -\langle\hat{x}_2\rangle\langle\hat{p}_2\rangle & \langle\hat{p}_2^2\rangle - \langle\hat{p}_2\rangle^2\\
\end{array} \right] \; , $}
\label{Bmat}
\end{equation}
\begin{equation}
 C=\resizebox{.9\hsize}{!}{$
\left[ 
 \begin{array}{cc}
 \langle \hat{x}_1\hat{x}_2 + \hat{x}_2\hat{x}_1\rangle/2 - \langle\hat{x}_1\rangle \langle\hat{x}_2\rangle      & \langle \hat{x}_1\hat{p}_2 + \hat{p}_2\hat{x}_1\rangle /2 -\langle\hat{x}_1\rangle \langle\hat{p}_2\rangle\\
 \langle \hat{p}_1\hat{x}_2 + \hat{x}_2\hat{p}_1\rangle /2 -\langle\hat{x}_2\rangle \langle\hat{p}_1\rangle   &  \langle \hat{p}_1\hat{p}_2 + \hat{p}_2\hat{p}_1\rangle/2 -\langle\hat{p}_1\rangle \langle\hat{p}_2\rangle \\
\end{array} \right] \; , $}
\label{Cmat}
\end{equation}
$C$ denoting the correlation noise.

{\bf \small Entanglement measure.}--- A two-mode Gaussian state is separable if the symplectic eigenvalues of the partial transposed state satisfy
\begin{equation}
2\nu_- > 1 \quad \text{or} \quad 2\nu_{+} > 1 \; .
\label{etacond}
\end{equation}
Symplectic eigenvalues can be calculated from
\begin{equation}
\nu_{\pm} \equiv \frac{1}{\sqrt{2}} \left( \sigma(V)  \pm \left\{  [\sigma(V)]^2-4{\rm det}(V)   \right\}^{1/2}   \right)^{1/2} \; ,
\label{eta}
\end{equation}
where $\sigma(V)={\rm det}(A)+{\rm det}(B)-2{\rm det}(C)$. Since $\nu_-<\nu_+$, one needs to check if $2\nu_- <1$ for the presence of the entanglement between the two modes. $2\nu_-<1$ is both a necessary and a sufficient criterion for Gaussian states. However much the $2\nu_-$ is smaller than $1$, that much stronger the entanglement is. Hence, one calculates the logarithmic negativity~(log-neg)
\begin{equation}
E_{\cal N}={\rm max}(0,-{\rm \log}(2\nu_-)) \; ,
\label{EN}
\end{equation}
an entanglement measure (monotone) for Gaussian states~\cite{Plenio2005}.  

$E_{\cal N}$ is zero if $2\nu^- >1$ and increases as symplectic eigenvalue becomes $2\nu^- < 1$.

{\bf \small Entanglement criterion.}--- Logarithmic negativity is a quantity which can witness the entanglement in Gaussian states~\cite{AdessoPRA2004,WernerPRA2002}. Because, Williamson theorem, used in the derivation of $E_{\cal N}$, valid for Gaussian states. The two-mode entanglement criterion
\begin{eqnarray}
\lambda_{\rm \scriptscriptstyle Simon}&&= {\rm det}(A){\rm det}(B) + \left(\frac{1}{4}-|{\rm det}(C)|\right)^2 \nonumber
\\
&&- {\rm tr}(AJCJBJC^TJ) \nonumber  \\
&&- \frac{1}{4} \left( {\rm det}(A) + {\rm det}(B) \right) \quad  < \quad 0 ,
\label{lambdaSimon}
\end{eqnarray}
Simon derives in Ref.~\cite{SimonPRL2000}, on the other hand, can be used to witness the two-mode entanglement for non-Gaussian states, too. Although a negative value of $\lambda_{\rm \scriptscriptstyle Simon}$ is ``sufficient'' to witness the presence of entanglement for any quantum state, it is natural to expect $\lambda_{\rm \scriptscriptstyle Simon}$ to be successful when the nonclassicality (entanglement) is associated with a quadrature type squeezing, e.g. of mixed operators $\alpha \hat{x}_1 + \beta \hat{x}_2$~\cite{Duan&Zoller2000}.

The criterion Simon derives~\cite{SimonPRL2000}, and we state above, is particularly in a strong form. Because $\lambda_{\rm \scriptscriptstyle Simon}$ is written in local-invariant terms, e.g. invariant under local intra-mode rotations $\hat{a}_{1,2}(\phi_{1,2})=e^{i\phi_{1,2}} \hat{a}_{1,2}$. SPH criterion, too, is also a necessary and sufficient criterion for Gaussian states, while it is not demonstrated as a measure unlike log-neg. Nevertheless, $\lambda_{\rm \scriptscriptstyle Simon}$ is valuable since it can be used to witness the entanglement of quantum states other than Gaussian states. The $J$ matrix is
\begin{equation}
J=\left[ 
 \begin{array}{cc}
 0   & 1 \\
 -1  &  0 \\
\end{array} \right] \; . 
\end{equation}

%%%%%%%%%%%%%%%%%%%%%%%%%%%%%%%%%%%%%%%%%%%%%%%%%%%%%%%%%%%%%%%%%%%%%%%%%%%%%%%%%%%%%%%%%%%%%%%%%%%%%%%%%%%
\subsection{Single-mode nonclassicality} \label{sec:SMNc}

{\bf \small Beam splitter approach}--- A single-mode field $\hat{a}$, mixed with a coherent (or in particular a vacuum) state, generates the two output modes, $\hat{a}_1$ and $\hat{a}_2$, at the beam splitter output. The nonclassicality of the $\hat{a}$-mode is quantified by the strength of the entanglement it can generate with linear beam-splitter operations~\cite{EntanglementPot}.

Using this argument, it is possible to (i) measure the single-mode nonclassicality of a Gaussian input state by counting the maximum entanglement the single-mode nonclassical state can generate at the beam-splitter output. It is also possible to (ii) reveal if an input state is single-mode nonclassical by checking if the two output modes are entangled using $\lambda_{\scriptscriptstyle \rm Simon}$.

Here, we express both (i) the measure for single-mode Gaussian states and (ii) the nonclassicality condition for a single-mode state in terms of $\langle\hat{a}^2\rangle$ and $\langle\hat{a}^\dagger\hat{a}\rangle$. $\hat{a}$ is the annihilation operator for the single-mode input state whose single-mode nonclassicality we aim to measure/witness. 

The methodology is as follows. ($\tt 1$) We evaluate the variance terms, like $\langle\hat{a}_{1,2}^2\rangle$ and $\langle\hat{a}_1\hat{a}_2\rangle$, in terms of $\langle\hat{a}^2\rangle$ and $\langle\hat{a}^\dagger\hat{a}\rangle$ using the beam-splitter transformations, see Eqs.~(\ref{BStransa}) and (\ref{BStransb}). Next, we evaluate the elements of the covariance matrix, Eqs.~(\ref{Vmatrix}-\ref{Cmat}). This way, the covariance matrix $V_{ij}$ becomes dependent only on the $\langle\hat{a}^2\rangle$ and $\langle\hat{a}^\dagger\hat{a}\rangle$ values, and the transmission ($t$), reflection ($r$) and phase ($\phi$) parameters of the beam-splitter. ($\tt 3$) Finally, we calculate the $E_{\cal N}$ from $V_{ij}$ matrix elements using Eqs. (\ref{eta}) and (\ref{EN}) in order to obtain the (i) single-mode nonclassicality measure ${\cal N}_{\rm \scriptscriptstyle SMNc}$. We also calculate $\lambda_{\rm \scriptscriptstyle Simon}$ to obtain a single-mode nonclassicality (ii) condition $\eta_{\rm \scriptscriptstyle SMNc}<0$. In order to calculate the maximum two-mode entanglement at the beam splitter output, we maximize $E_{\cal N}(t,\phi)$ with respect to $t=[0,1]$ and $\phi=[0,2\pi]$, with $r^2+t^2=1$. We refer the maximum value of $E_{\cal N}(t,\phi)$ as ${\cal N}_{\rm \scriptscriptstyle SMNc}$. Similarly, we minimize $\lambda_{\rm \scriptscriptstyle Simon}(t,\phi)$ with respect to beam splitter parameters. That is, we search for the smallest value of $\lambda_{\rm \scriptscriptstyle Simon}(t,\phi)$ which we refer as $\eta_{\rm \scriptscriptstyle SMNc}$. A negative value of $\eta_{\rm \scriptscriptstyle SMNc}$ witnesses the nonclassicality of the input state~\cite{Kim&Knight2002}.

The two-mode state, generated at the output of a beam-splitter from a single-mode input state $|\psi_a\rangle=\sum_{n=0}^{\infty} c_n |n\rangle$, can be calculated \cite{Kim&Knight2002,BSentanglement,aharanov1966} using the transformed operators
\begin{subequations}
\begin{eqnarray}
\hat{a}_1(\xi)=\hat{B}^\dagger(\xi)\hat{a}_1\hat{B}(\xi)=te^{i\phi}\hat{a}_1 + r\hat{a}_2 \; ,
\label{BStransa}
\\
\hat{a}_2(\xi)=\hat{B}^\dagger(\xi)\hat{a}_2\hat{B}(\xi)=-r\hat{a}_1 + te^{-i\phi}\hat{a}_2 \; ,
\label{BStransb}
\end{eqnarray}
\end{subequations}
where 
\begin{equation}
\hat{B}(\xi)=e^{\xi\hat{a}_2^\dagger \hat{a}_1 - \xi^*\hat{a}_1^\dagger \hat{a}_2} 
\end{equation}
is the beam-splitter operator. In calculating the expectation values, one uses $|\psi_a\rangle_1\otimes|0\rangle_2$ as the initial state of the two-mode output, where input state $|\psi_a\rangle$ is placed into the state of the first mode ($\hat{a}_1$). This is equivalent to Schr\"{o}dinger picture \cite{aharanov1966} where two-mode state is determined as $|\psi_{12}\rangle = f(\mu_1 \hat{a}_1^\dagger + \mu_2 \hat{a}_2^\dagger) |0\rangle_1\otimes|0\rangle_2 $  with $f(\hat{a}^\dagger)$ is the expansion of the input mode $|\psi_a\rangle=\left( \sum_{n=0}^\infty  d_n (\hat{a}^\dagger)^2\right) |0\rangle_a$, that is $f(\hat{a}^\dagger)=\sum_{n=0}^{\infty} d_n (\hat{a}^\dagger)^n$.

As an explicit example, working in the Heisenberg picture, one can calculate the expectation value as
\begin{equation}
\langle\hat{a}_1\hat{a}_2\rangle = {}_2\langle0| \otimes {}_1\langle \psi_a| \hat{B}^\dagger(\xi) \hat{a}_1\hat{a}_2 \hat{B}(\xi) |\psi_a\rangle_1\otimes |0\rangle_2,
\end{equation}
which is equal to 
\begin{eqnarray}
\langle\hat{a}_1\hat{a}_2\rangle = \nonumber \hspace{2.5in}
\\ 
 {}_2\langle0| \otimes {}_1\langle \psi_a|  (te^{i\phi}\hat{a}_1+r\hat{a}_2) (-r\hat{a}_1+te^{-i\phi}\hat{a}_2)
 |\psi_a\rangle_1\otimes |0\rangle_2 \nonumber
 \\
\quad
\label{a1a2explicit}
\end{eqnarray}
using the transformations (\ref{BStransa}) and (\ref{BStransb}). Since ${}_1\langle \psi_a|\hat{a}_1^2|\psi_a\rangle_1=\langle \psi_a|\hat{a}^2|\psi_a\rangle\equiv\langle\hat{a}^2\rangle$, Eq.~(\ref{a1a2explicit}) becomes
\begin{equation}
\langle\hat{a}_1\hat{a}_2\rangle = -tre^{i\phi} \langle\hat{a}^2\rangle \; .
\end{equation}
Similarly, other matrix elements of $V_{ij}$ can be calculated as
\begin{widetext}
\begin{equation}
 A=\left[  \begin{array}{cc}
t^2 [ \cos(\theta + 2\phi)\: v_a+n_a ] + \frac{1}{2}  &  t^2v_a\sin(\theta+2\phi)\\
 t^2v_a\sin(\theta+2\phi)&  t^2 [ -\cos(\theta + 2\phi)\: v_a+n_a ] + \frac{1}{2}  \\
\end{array} \right] \; , 
\label{Amatsinglemode}
\end{equation}
\begin{equation}
 B=\left[ 
 \begin{array}{cc}
r^2(\cos\theta\: v_a + n_a)+\frac{1}{2} & r^2v_a\sin\theta \\
r^2v_a\sin\theta & r^2(-\cos\theta\: v_a + n_a)+\frac{1}{2} \\
\end{array} \right] \; , 
\label{Bmatsinglemode}
\end{equation}
\begin{equation}
 C=tr\left[  \begin{array}{cc}
- [ \cos(\theta + \phi)v_a+ \cos\phi\: n_a ]   &  [- \sin(\theta + \phi)v_a+ \sin\phi\: n_a ]   \\
-[ \sin(\theta + \phi)v_a+ \sin\phi\: n_a ] &   [ \cos(\theta + \phi)v_a - \cos\phi \: n_a ]   \\
\end{array} \right] \; , 
\label{Cmatsinglemode}
\end{equation}
\end{widetext}
where $\langle\hat{a}^2\rangle=v_ae^{i\theta}$ with $v_a$ is real and positive, and $\langle\hat{a}^\dagger\hat{a}\rangle=n_a$~\footnote{In the experiments one measures $\langle(\delta\hat{x})^2\rangle$ and $\langle(\delta\hat{p})^2\rangle$ to obtain $\langle(\delta\hat{a})^2\rangle$ and $\langle\delta\hat{a}^\dagger\delta\hat{a}\rangle$}. We remark that in the derivation of Eqs.~(\ref{Amatsinglemode}-\ref{Cmatsinglemode}) we applied a linear transformation which makes the expectations $\langle \hat{x}_i\rangle=\langle \hat{p}_i\rangle=0$, or simply $\langle\hat{a}_i\rangle=0$. Such a transformation does not affect the noise features, i.e. $\delta \hat{a}_i=\hat{a}-\langle \hat{a}_i\rangle$, with which nonclassicality features are determined in the $V$ matrix~\cite{SimonPRL2000,simonPRA1994}. In other words, $n_a$ and $v_a$ in Eqs.~(\ref{Amatsinglemode}-\ref{Cmatsinglemode}) are $\langle\delta\hat{a}^\dagger \delta\hat{a}\rangle$ and $\langle(\delta\hat{a})^2\rangle$, respectively. Here, $\delta\hat{a}$, $\hat{a}=\langle\hat{a}\rangle + \delta \hat{a}$, is the noise operator with zero expectation $\langle \delta\hat{a}\rangle=0$ \cite{Vitali2008PRA}.

{\bf \small Single-mode nonclassicality measure for Gaussian states.}--- Therefore, given the $\langle\hat{a}^2\rangle$ and $\langle\hat{a}^\dagger\hat{a}\rangle$ for a single-mode state, one can calculate the degreee of two-mode entanglement $E_{\cal N}$ this input mode generates through a beam-splitter. One inserts Eqs.~(\ref{Amatsinglemode})-(\ref{Cmatsinglemode}) into Eq.~(\ref{eta}) to determine $\nu_-$ and uses Eq.~(\ref{EN}) to obtain the measure of entanglement $E_{\cal N}$ at the output. This, calculated measure ($E_{\cal N}$), is also the degree of nonclassicality of the single-mode sate $|\psi_a\rangle$ \cite{Vogel&Sperling2014,Mraz&Hage2014}. After a maximization of $E_{\cal N}(t,\phi)$ over $t$ and $\phi$ is carried out, with reflection $r^2=1-t^2$ is already constrained, we obtain the single-mode nonclassicality measure ${\cal N}_{\rm \scriptscriptstyle SMNc}$.

{\bf \small Single-mode nonclassicality condition.}--- When the two-mode entanglement criterion ${\lambda}_{\rm \scriptscriptstyle Simon}(t,\phi)$ is calculated from Eqs.~(\ref{Amatsinglemode}-\ref{Cmatsinglemode}) via Eq.~(\ref{lambdaSimon}), ${\lambda}_{\rm \scriptscriptstyle Simon}(t,\phi)<0$ becomes a single-mode nonclassicality condition. We define $\eta_{\rm \scriptscriptstyle SMNc}$ as the minimized value of ${\lambda}_{\rm \scriptscriptstyle Simon}(t,\phi)$ over $t$ and $\phi$.

%%%%%%%%%%%%%%%%%%%%%%%%%%%%%%%%%%%%%%%%%%%%%%%%%%%%%%%%%%%%%%%%%%%%%%%%%%%%%%%%%%%%%%%%%%%%%%%%%%%%%%%%%%%
\section{Examples on the behavior of $E_{\cal N}$ and ${\cal N}_{\rm \scriptscriptstyle SMNc}$} \label{sec:examples}

In this section, we first measure the single-mode nonclassicality~(SMNc) of a quadrature-squeezed state. Second, we consider a second harmonic generating damped nonlinear cavity where vacuum noise works against the onset of nonclassicality. We find the pump strengths over which single-mode nonclassicality becomes resistant to the vacuum noise. Third, we consider a quantum superradiant phase transition where states are non-Gaussian. WE use the single-mode nonclassicality condition $\eta_{\rm \scriptscriptstyle SMNc}<0$ to witness the nonclassicality of the superradiatly emitted single mode light.

%%%%%%%%%%%%%%%%%%%%%%%%%%%%%%%%%%%%%%%%%%%%%%%%%%%%%%%%%%
\subsection{Single-mode squeezed state} \label{sec:squeezing}

A nonclassical squeezed state can be generated by applying the squeezing operator $\exp[\beta^*\hat{a}^2-\beta(\hat{a}^{\dagger})^2]$ to a coherent state $|\alpha\rangle$, in particular to vacuum $|0\rangle$. Here, $\beta=re^{i\theta}$ determines the strength ($r$) and the phase ($\theta$) of the quadrature squeezing. For a squeezed coherent state~\cite{Scullybook}
\begin{eqnarray}
\langle\hat{a}\rangle = {\rm C}\alpha-{\rm S}e^{i\theta}\alpha^* \; , \hspace{2in}
\\
\langle\hat{a}^2\rangle={\rm C}^2\alpha^2 {\rm S}^2e^{i2\theta} {\alpha^*}^2 -{\rm CS} e^{i\theta} (2|\alpha|^2+1) \; , \hspace{0.6 in}
\\
\langle\hat{a}^\dagger\hat{a}\rangle = {\rm C}^2 |\alpha|^2 + {\rm S}^2(1+|\alpha|^2) - {\rm CS}(e^{i\theta}{\alpha^*}^2 + e^{-i\theta}\alpha^2) \hspace{0.21in}
\end{eqnarray} 
where C$\equiv\cosh r$ and S$\equiv\sinh r$. The variances of the field, to use in Eqs.~(\ref{Amatsinglemode})-(\ref{Cmatsinglemode}), can be calculated from $\langle\delta\hat{a}^2\rangle=\langle\hat{a}^2\rangle - \langle\hat{a}\rangle^2$ and $\langle\delta\hat{a}^\dagger\delta\hat{a}\rangle=\langle\hat{a}^\dagger\hat{a}\rangle - |\langle\hat{a}\rangle|^2$.

\begin{figure}
\includegraphics[width=3.4in]{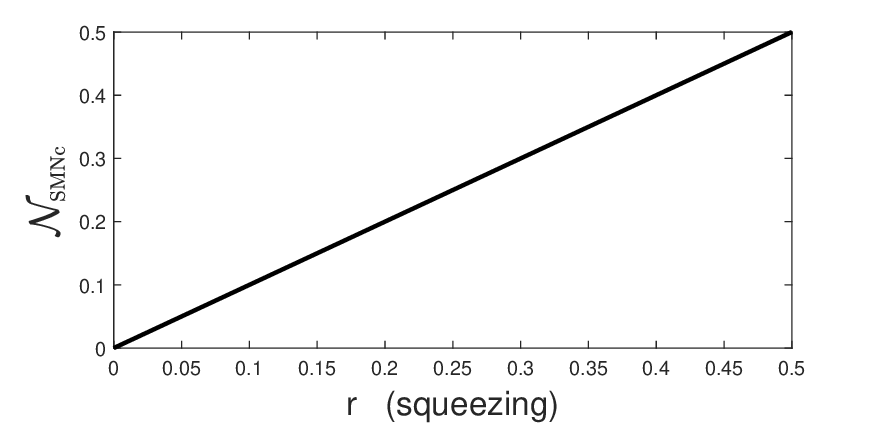}
\caption{Degree of nonclassicality ${\cal N}_{\rm \scriptscriptstyle SMNC}$, i.e. optimized value of  $E_{\cal N}$ with respect to beam splitter parameters, versus squeezing strength $r$, in a single-mode quadrature-squeezed state. We observe that ${\cal N}_{\rm \scriptscriptstyle SMNC}=r$.}
\label{fig1}
\end{figure}

In Fig.~\ref{fig1}, we plot the dependence of the degree of nonclassicality ${\cal N}_{\rm \scriptscriptstyle SMNC}$, i.e. the value$E_{\cal N}$ optimized with respect to beam splitter parameters, with respect to squeezing strength $r$. We observe that; not only ${\cal N}_{\rm \scriptscriptstyle SMNC}$ increases with squeezing $r$, but also the relation ${\cal N}_{\rm \scriptscriptstyle SMNC}=r$ holds for the amount of the single-mode nonclassicality.

%
%
%It is a well known phenomenon that emergence and the degree of two-mode entanglement depends on the choice of the phase of the parametrized pump (approximation $\langle\hat{a}_0\rangle \rightarrow \alpha_0=|\alpha_0|e^{i\theta_0}$), see Eq.~(13) in Ref.~\cite{metasginEndfire} and Refs.~\cite{parametricPhase1,parametricPhase2,parametricPhase3}. Hence, in Fig.~\ref{fig1} we give two plots. Solid-line is for a fixed beam splitter angle $\phi=0$. In the calculation of the dotted-line we search over $\phi$ values which maximizes the $E_{\cal }$ for a given squeezing strength $r$. We observe that by tuning the squeezing angle one can obtain larger degrees of nonclassicality.

%
%
%%%%%%%%%%%%%%%%%%%%%%%%%%%%%%%%%%%%%%%%%%%%%%%%%%%%%%%%%%
\subsection{Second harmonic generator crystal} \label{sec:SHG}

In this part, we consider a nonlinear crystal generating a second harmonic signal. We calculate the nonclassicality the second harmonic process generates in the first harmonic mode $\hat{a}_1$. We consider the quantum (noise) fluctuations $\delta\hat{a}_1$ around the steady-state values $\langle\hat{a}_1\rangle=\alpha_1$, i.e. $\langle\delta\hat{a}_1\rangle=0$~\cite{VitaliPRL2007}. We ignore higher-order noise terms like $\hat{a}_1\hat{a}_2$ and, thus, ensure that states remain Gaussian~\cite{Vitali2008PRA}. A detailed treatment of this standard method, e.g. for an optomechanical system, can be found in Refs.~\cite{Vitali2008PRA,VitaliPRL2007,QuantumNoiseBook}.

Modes of the cavity, relevant with the second harmonic process, are $\hat{a}_1$, of resonance $\Omega_1$ and excited with a strong pump laser $\sim e^{-i\omega t}$, and $\hat{a}_2$, of resonance $\Omega_2$ and into which second harmonic conversion (signal) $\sim e^{-i2\omega t}$ takes place. Both crystal modes are coupled to (damp into) the vacuum modes. The noise introduced due to vacuum fluctuation are treated within the input-output formalism~\cite{Vitali2008PRA,VitaliPRL2007,QuantumNoiseBook}.

The hamiltonian of such a system can be composed as
\begin{equation}
{\cal H}= \hbar\Delta_1 \hat{a}_1^\dagger \hat{a}_1 + \hbar\Delta_2 \hat{a}_2^\dagger \hat{a}_2 + i\hbar \varepsilon_{\rm \scriptscriptstyle L}(\hat{a}_1^\dagger-\hat{a}_1)
+ \hbar\chi (\hat{a}_2^\dagger \hat{a}_1^2 + \hat{a}_1^2{}^\dagger \hat{a}_2)
\end{equation}
in the frame rotating with $\omega$ and $2\omega$ for $\hat{a}_1$ and $\hat{a}_2$, respectively. The first two terms are the photon occupation energies of the two crystal modes $\hat{a}_{1,2}$ with rotating frame energies $\Delta_1=\Omega_1-\omega$ and $\Delta_2=\Omega_2 - 2\omega$. Here, $\omega$ is the frequency of the pump laser. In the third term, the laser, of amplitude $\sim \varepsilon_{\rm \scriptscriptstyle L}$, pumps the cavity mode $\hat{a}_1$. More explicitly, it is $\varepsilon_{\rm \scriptscriptstyle L}=g_1 \alpha_{\rm \scriptscriptstyle L}$, with $g_1$ is the coupling of the $\hat{a}_1$ mode to the vacuum and $|\alpha_{\rm \scriptscriptstyle L}|^2$ is the proportional to the number of laser photons. In the last term, there happens the second harmonic conversion process. Two $\hat{a}_1$ crystal mode photons, each oscillating as $e^{-i\omega t}$, combine to generate a single $\hat{a}_2$ crystal mode photon oscillating as $e^{-i2\omega t}$. $\chi$ is an overlap integral between two cavity modes, i.e. between eigenmodes $u_1^2({\bf r})$ and $u_2({\bf r})$, and it is also proportional to the second harmonic nonlinear response of the crystal. For simplicity, we do not consider experimental complexities like phase-matching etc. 

We examine the noise properties of the pumped $\hat{a}_1$ mode, i.e. $\delta\hat{a}_1$. First, we obtain the Langevin equations
\begin{eqnarray}
&&\dot{\hat{a}}_1 = -(i\Delta_1+\gamma_1)\hat{a}_1 - i2\chi \hat{a}_1^\dagger \hat{a}_2 + \varepsilon_{\rm \scriptscriptstyle L}, \label{Langa1}
\\
&&\dot{\hat{a}}_2 = -(i\Delta_2+\gamma_2)\hat{a}_2 - i2\chi \hat{a}_1^2, \label{Langa2}
\end{eqnarray}
where $\gamma_{1,2}$ are the damping of the $\hat{a}_{1,2}$ mode into the vacuum. The Langevin equations for the noise fluctuations $\delta \hat{a}_{1,2}$, over the steady-state values $\alpha_{1,2}$, i.e. $\hat{a}_{1,2}=\alpha_{1,2}+\delta \hat{a}_{1,2}$, can be obtained as
\begin{equation}
\delta\dot{\hat{a}}_1 = -(i\Delta_1+\gamma_1)\delta\hat{a}_1 - i2\chi (\alpha_1^* \delta \hat{a}_2 + \alpha_2\delta\hat{a}_1^\dagger) + g_1 \delta \hat{a}_{1,{\rm in}}(t),\label{dela1}
\end{equation}
\begin{equation}
\delta\dot{\hat{a}}_2 = -(i\Delta_2+\gamma_2)\delta\hat{a}_2 - i2\chi \alpha_1 \delta\hat{a}_1 + g_2 \delta \hat{a}_{2,{\rm in}}(t), \label{dela2}
\end{equation}
where we ignore the second-order noise terms like $\delta\hat{a}_1 \delta\hat{a}_2$~\cite{Vitali2008PRA,VitaliPRL2007,QuantumNoiseBook}. Steady-state values $\alpha_{1,2}$ are determined from Eqs.~(\ref{Langa1},\ref{Langa2}).  $ g_{i} \delta \hat{a}_{i, {\rm in}}(t)=-i\sum_{{\bf k}} e^{-i\omega_k t} \hat{b}_{\bf k}(0)$ represents the input vacuum noise, where $\hat{b}_{\bf k}(0)$ are the vacuum operators~\cite{Vitali2008PRA,VitaliPRL2007,QuantumNoiseBook}.  This treatment, linearization of the noise operators, ensures that quantum states stay Gaussian~\cite{Vitali2008PRA}. $g_{1,2}$ are the coupling of the crystal modes to vacuum and they are related to the damping rates as $\gamma_{1,2}=\pi D(\omega_{1,2}) g_{1,2}^2$~\cite{Scullybook} where $D(\omega_{1,2})$ is the density of states at $\omega_{1,2}$~\footnote{We note that final results necessitate the knowledge of $\gamma_{1,2}$ only. For this reason, $g_{1,2}$ are usually referred as $\sqrt{\gamma_{1,2}}$ in most treatments~\cite{Vitali2008PRA,VitaliPRL2007,QuantumNoiseBook}.}.

Eqs.~(\ref{dela1}) and (\ref{dela2}) are linear and can be solved exactly. Calculations can be found in Refs.~\cite{Vitali2008PRA,VitaliPRL2007,QuantumNoiseBook}. We calculate $v_a=\langle\delta\hat{a}_1^2\rangle$ and $n_a=\langle\delta\hat{a}_1^\dagger \delta\hat{a}_1\rangle$ , and place into Eqs.~(\ref{Amatsinglemode}-\ref{Cmatsinglemode}) in order to evaluate the nonclassicality ${\cal N}_{\rm \scriptscriptstyle SMNc}$ the second harmonic process generates  in the $\hat{a}_1$ crystal mode. We also check if the Eqs.~(\ref{dela1}) and (\ref{dela2}) are stable and we determine an $\varepsilon_{\rm \scriptscriptstyle L}^{\rm (crt)}$, for $\varepsilon_{\rm \scriptscriptstyle L}<\varepsilon_{\rm \scriptscriptstyle L}^{\rm (crt)}$ the solutions for noise fluctuations are stable.

We note that, our aim in this paper is not to present details for such a second harmonic calculations, but it is rather to examine the nonclassicality measure on a toy system which is related to a nonlinear frequency conversion process.

\begin{figure}
\includegraphics[width=3.4in]{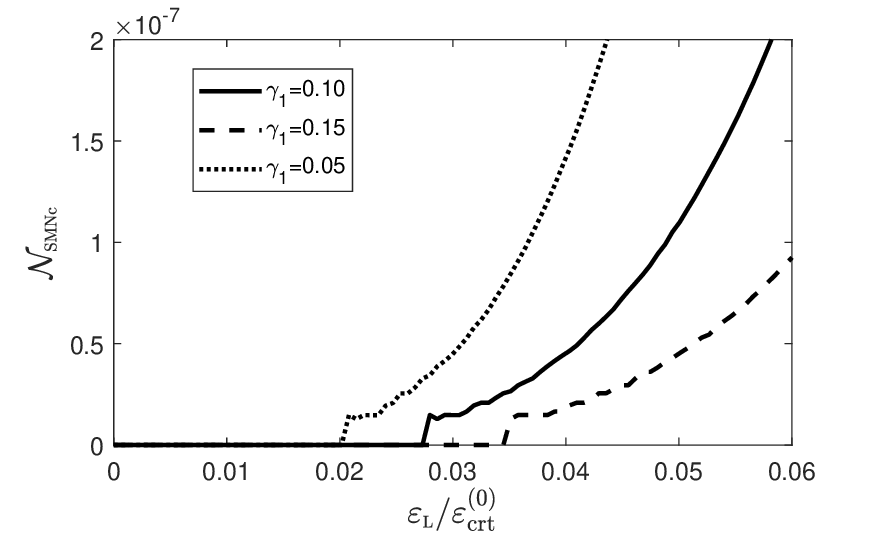}
\caption{Onset of single-mode nonclassicality in a damped nonlinear crystal generating second harmonic conversion. The nonclassicality  ${\cal N}_{\rm \scriptscriptstyle SMNc}$ of the $\hat{a}_1$ mode is investigated for the second harmonic process $(\hat{a}_2^\dagger \hat{a}_1^2 + \hat{a}_1^2{}^\dagger \hat{a}_2)$. Laser strength $\varepsilon_{\rm \scriptscriptstyle L}$ is scaled with a common $\varepsilon_{\rm \scriptscriptstyle L}^{\rm (crt)}\equiv \varepsilon_{\rm crt}^{\rm (0)} $ which belongs to the damping $\gamma_1=0.1$. For a larger damping, nonclassicality onsets (becomes resistant against the vacuum noises) for a stronger second harmonic process, as could be expected intuitively.  }
\label{fig2}
\end{figure}

In Fig.~\ref{fig2}, we evaluate $E_{\cal N}(\langle\delta\hat{a}_1^2\rangle, \langle \delta\hat{a}_1^\dagger \delta \hat{a}_1\rangle)$ for the $\hat{a}_1$ mode and present the optimized single-mode nonclassicality~(SMNc) measure ${\cal N}_{\rm \scriptscriptstyle SMNc}$. We choose the parameters $\Delta_{1,2}=1$, $\chi=0.001$, $\gamma_{1,2}=0.1$, i.e. scaled with the detuning $\Delta_1=1$. In Fig.~\ref{fig2}, we scale the laser pump strength $\varepsilon_{\rm \scriptscriptstyle L}$ with the critical value $\varepsilon_{\rm \scriptscriptstyle L}^{\rm (crt)}$ over which the solutions for noise operators become unstable.

\begin{figure}
\includegraphics[width=3.4in]{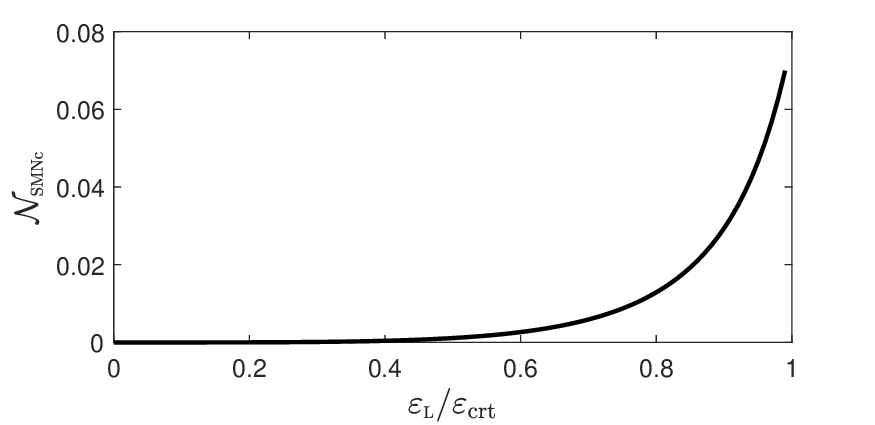}
\caption{Single-mode nonclassicality of $\hat{a}_1$ mode in the second harmonic process taking place in the damped cavity. ${\cal N}_{\rm \scriptscriptstyle SMNc}$ reaches values as high as 0.07 before the instability takes place. This value can be compared with Fig.~\ref{fig1}, where squeezing hamiltonian is in action with a squeezing parameter $r=$0.07, but, without any damping. }
\label{fig3}
\end{figure}

We observe that upto a strength for the second harmonic process ---i.e. laser pump strength $\varepsilon_{\rm \scriptscriptstyle L}$, since $\chi$ is fixed--- the nonclassicality of the $\hat{a}_1$ crystal mode is destroyed by the vacuum noise, i.e. ${\cal N}_{\rm \scriptscriptstyle SMNc}=0$. For the stronger pump strength, e.g for $\varepsilon_{\rm \scriptscriptstyle L}/\varepsilon_{\rm \scriptscriptstyle L}^{\rm (crt)}>$0.02, a single-mode nonclassicality onsets in the $\hat{a}_1$ mode. In Fig.~\ref{fig2} we also observe that, for a higher cavity damping $\gamma_1$, onset of single-mode nonclassicality of the $\hat{a}_1$ mode takes place a higher pump strength, as one could expect intuitively. In Fig.~\ref{fig3}, we also present the nonclassicality ${\cal N}_{\rm \scriptscriptstyle SMNc}$ for higher pump strengths, upto a critical strength over which solutions for noise operators become unstable. The maximum value the second harmonic generation process, for given parameters, is ${\cal N}_{\rm \scriptscriptstyle SMNc}=$0.07 which corresponds to a squeezing parameter of $r=0.07$ in Fig.~\ref{fig1}. One should, however, note that in Fig.~\ref{fig1} we do not consider any damping.

%

%
%%%%%%%%%%%%%%%%%%%%%%%%%%%%%%%%%%%%%%%%%%%%%%%%%%%%%%%%%%
\subsection{Superradiant phase-transition} \label{sec:Superradiance}

When an ensemble of 2-level atoms are pumped, above a critical intensity the coupling of matter and light induces a new phase. This is called as Dicke phase-transition \cite{Emary&BrandesPRE2003,KetterleSience1999,Skribanowitz1973,Dicke1954} or superradiant phase where all atoms radiate collectively. Such a phase-transition is accompanied by a jump in the bipartite entanglement among the constituent atoms~\cite{BrandesPRL2004}, while violation of many-particle entanglement criteria accompanies the strength of the cooperation in the superradiant phase~\cite{tasgin2017many,BrandesPRA2005Q}.

The Hamiltonian for this system can be written as \cite{Emary&BrandesPRE2003}
\begin{equation}
\hat{\cal H}=\hbar\omega\hat{a}^\dagger\hat{a} + \hbar\omega_{eg}\hat{S}_z + \hbar\frac{g}{\sqrt{N}} (\hat{S}_+\hat{a} + \hat{S}_-\hat{a}^\dagger) \; ,
\label{HSR}
\end{equation}
where $\hat{S}_\pm$ and $\hat{S}_z$ are the collective raising/lowering and energy level operators for the $N$ identical 2-level atom system~\cite{Emary&BrandesPRE2003}. $\omega_{eg}$ is the level spacing of a single 2-level atom. $\omega$ is the frequency of the superradiantly emitted light and $\hat{a}$ is the annihilation operator of this single-mode field. $g$ is the interaction strength between a single atom and photon. The superradiant phase transition occurs above $g>g_c=\sqrt{\omega_{eg}\omega}$~\cite{Emary&BrandesPRE2003}.

Similar to Refs.~\cite{BrandesPRL2004,tasgin2017many}, we numerically evaluate the ground state of the hamiltonian~(\ref{HSR}) by choosing $N=$80 identical atoms and 142 dimensions for the field mode $\hat{a}$. Since the particles are identical, the dimension of the atomic system is only $2S+1=N+1=81$, i.e. not $2^N$. Because, atoms can occupy only the symmetric Dicke states~\cite{symmetricDicke}, i.e. $|\psi\rangle=\sum_m c_m |S=N/2,m\rangle$ or $|\psi\rangle = \sum_{n_e} |n_g,n_e\rangle$ with $n_g+n_e=N$.

\begin{figure}
\includegraphics[width=3.4in]{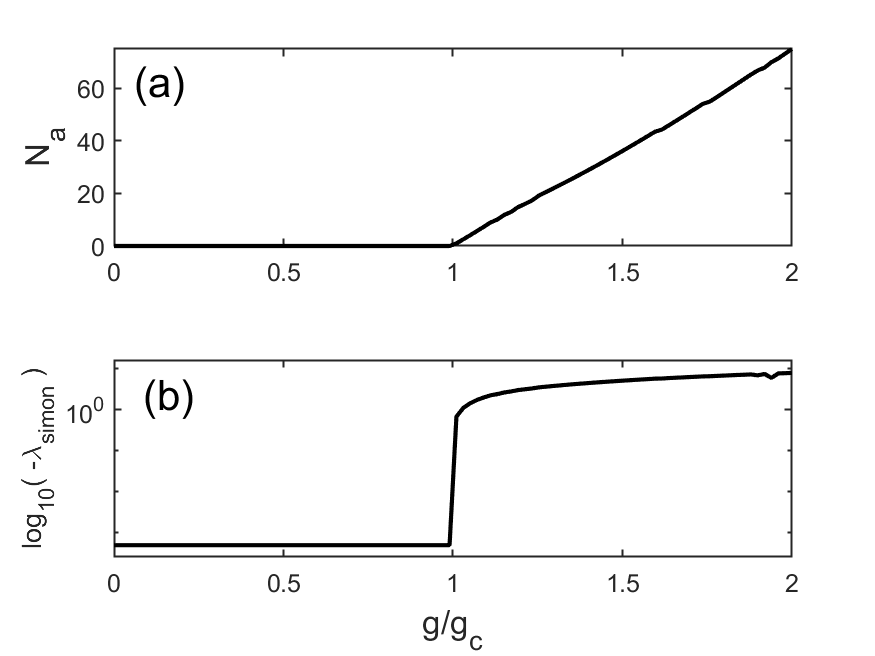}
\caption{A superradiant phase transition in an atomic ensemble, coupled with a single-mode field, see Eq.~(\ref{HSR}). (a) Mean photon occupation $N_a$, in the ground state of the interacting system (\ref{HSR}), becomes macroscopic~\cite{Emary&BrandesPRE2003} above a critical light-matter coupling $g>g_c$. (b) Violation of the single-mode nonclassicality condition, i.e. negativity of $\eta_{\rm \scriptscriptstyle SMNc}$, accompanies the strength of the superradiant phase.}
\label{fig4}
\end{figure}

In Fig.~\ref{fig4}a, we plot the mean photon occupation of the ground state of the hamiltonian (\ref{HSR}) with respect to the atom-photon coupling strength $g$. We observe that for $g<g_c$, both atoms and field energetically favor to stay in the not-excited state. For $g>g_c$, the last term in (\ref{HSR}) becomes macroscopic, atoms move to excited state and photon occupancy emerges in the single-mode field $\hat{a}$ \cite{Emary&BrandesPRE2003}.

Superradiant states are not Gaussian. Thus, we cannot use the single-mode nonclassicality measure (also a condition) ${\cal N}_{\rm \scriptscriptstyle SMNc}$ which is derived for Gaussian states. The single-mode nonclassicality condition $\eta_{\rm \scriptscriptstyle SMNc}<0$, however, can be used for non-Gaussian quantum states~\footnote{In order to remove the possibility of a misunderstanding we note that, $\eta_{\rm \scriptscriptstyle SMNc}$ is only a sufficient conditions for quantum states other than Gaussian ones. Its violation, i.e. $\eta_{\rm \scriptscriptstyle SMNc}<0$, can witness the presence of single-mode nonclassicality, since it is obtained from $\lambda_{\rm \scriptscriptstyle Simon}$, while its positive values do not imply the absence of single-mode nonclassicality.}, too. 

Fig.~\ref{fig4}b shows that $\eta_{\rm \scriptscriptstyle SMNc}$ witnesses the presence of the single-mode nonclassicality in the superradiantly emitted light, into mode $\hat{a}$. Moreover, the strength of the violation of violation of $\eta_{\rm \scriptscriptstyle SMNc}$, i.e. its negativity, not a measure, accompanies the strength of the superradiant phase, similar to the many-particle entanglement in Ref.~\cite{tasgin2017many}. We plot the absolute value of $\lambda_{\rm \scriptscriptstyle Simon}$ in logarithmic scale for a better visualization.

%%%%%%%%%%%%%%%%%%%%%%%%%%%%%%%%%%%%%%%%%%%%%%%%%%%%%%%%%%%%%%%%%%%%%%%%%%%%%%%%%%%%%%%%%%%%%%%%%%%%%%%%%%%
%%%%%%%%%%%%%%%%%%%%%%%%%%%%%%%%%%%%%%%%%%%%%%%%%%%%%%%%%%%%%%%%%%%%%%%%%%%%%%%%%%%%%%%%%%%%%%%%%%%%%%%%%%%
%%%%%%%%%%%%%%%%%%%%%%%%%%%%%%%%%%%%%%%%%%%%%%%%%%%%%%%%%%%%%%%%%%%%%%%%%%%%%%%%%%%%%%%%%%%%%%%%%%%%%%%%%%%
\section{Nonclassicality condition from DGCZ criterion} \label{sec:SMNcfromDGCZ}

In this section, we demonstrate that the well-known Duan-Giedke-Cirac-Zoller~(DGCZ) criterion~\cite{Duan&Zoller2000} for two-mode entanglement, i.e. $\lambda_{\rm \scriptscriptstyle DGCZ}<0$, can be transformed into a single-mode nonclassicality condition $\eta_{\rm \scriptscriptstyle SMNc}^{\rm \scriptscriptstyle (DGCZ)}<0$. Moreover, when an optimization is conducted over $\lambda_{\rm \scriptscriptstyle DGCZ}$, the $\eta_{\rm \scriptscriptstyle SMNc}^{\rm \scriptscriptstyle (DGCZ)}$ turns out to be the generalized quadrature-squeezing condition $\langle \hat{a}^\dagger \hat{a}\rangle < |\langle \hat{a}^2\rangle| $. This condition does not necessitate an optimum choice (a search) for the direction of quadrature-squeezing, i.e. $\hat{a}_\phi=e^{i\phi} \hat{a}$.

 Duan-Giedke-Cirac-Zoller~(DGCZ) provide a criterion for detecting the presence of the entanglement in a two-mode system~\cite{Duan&Zoller2000}. This criterion is both  a necessary and a sufficient for two-mode Gaussian  states, similar to Simon-Peres-Horodecki~(SPH) criterion~\cite{SimonPRL2000}. A two-mode system is inseparable if the variance satisfies
\begin{equation}
\lambda_{\rm \scriptscriptstyle DGCZ}= \langle(\Delta\hat{u})^2\rangle + \langle(\Delta\hat{v})^2\rangle - \left( c^2+\frac{1}{c^2}\right) \; < \; 0 \; ,
\label{DGCZ1}
\end{equation}    
where $\hat{u}$ and $\hat{v}$ operators are defined as
\begin{eqnarray}
\hat{u}=|c|\hat{x}_1 + \hat{x}_2/c ,
\\
\hat{v}=|c|\hat{p}_1 - \hat{p}_2/c ,
\end{eqnarray}
with $c$ is a real number whose sign and magnitude will be fixed to ones which ``{\it minimize} $\lambda_{\rm DGCZ}(c)$". The criterion (\ref{DGCZ1}) can also be reformulated~\cite{metasginEndfire} as
\begin{eqnarray}
\lambda_{\rm \scriptscriptstyle DGCZ}= 2|c|^2\langle\hat{a}_1^\dagger\hat{a_1}\rangle + \frac{2}{|c|^2} \langle\hat{a}_1^\dagger\hat{a_1}\rangle \hspace{0.8in}
\nonumber
\\
+ \; 2{\rm sign}(c)\:  {\rm Re}\{ \langle\hat{a}_1\hat{a}_2\rangle + \langle\hat{a}_2\hat{a}_1\rangle \}  \; < \; 0 
\label{DGCZ2}
\end{eqnarray}
omitting first-order moments, again. Examining Eq.~(\ref{DGCZ2}), it becomes explicit that the magnitude of $c$ minimizing $\lambda_{\rm \scriptscriptstyle DGCZ}(c)$ is $c_*^2=\left( \langle\hat{a}_2^\dagger\hat{a}_2\rangle/\langle\hat{a}_1^\dagger\hat{a}_1\rangle \right)^{1/2}$ and sign of $c$ must be chosen opposite to ${\rm Re}\{ \langle\hat{a}_1\hat{a}_2\rangle + \langle\hat{a}_2\hat{a}_1\rangle \}$~\cite{metasginEndfire}.

We apply the beam splitter approach to obtain a necessary and sufficient condition for the nonclassicality of Gaussian single-mode states, from the Duan-Giedke-Cirac-Zoller criterion given in Eq.~(\ref{DGCZ2}). Using the beam splitter transformations \cite{Kim&Knight2002,BSentanglement,aharanov1966}, given in Eqs. (\ref{BStransa}) and (\ref{BStransb}),  one can obtain the expectation values as
\begin{eqnarray}
\langle\hat{a}_1^\dagger\hat{a}_1\rangle=t^2\langle\hat{a}^\dagger\hat{a}\rangle \quad {\rm and} \quad
\langle\hat{a}_2^\dagger\hat{a}_2\rangle=r^2\langle\hat{a}^\dagger\hat{a}\rangle \; ,
\\
\langle\hat{a}_1\hat{a}_2\rangle=\langle\hat{a}_2\hat{a}_1\rangle=-tre^{i\phi}\langle\hat{a}^2\rangle \; .
\end{eqnarray}
Inserting $c_*$ to obtain the minimum $\lambda_{\rm \scriptscriptstyle Simon}$, the nonclassicaity condition becomes
\begin{equation}
\lambda_{\rm \scriptscriptstyle Simon}=\langle\hat{a}^\dagger\hat{a}\rangle - {\rm sign}(c){\rm Re}\{ e^{i\phi} \langle\hat{a}^2\rangle \}.
\label{DGCZcond1}
\end{equation}
Eq.~(\ref{DGCZcond1}), actually, is also the quadrature-squeezing condition for $\hat{a}_\phi=e^{i\bar{\phi}} \hat{a}$, where $\hat{\phi}$ is the angle of rotation in the $x$-$p$ plane~\cite{Scullybook} for determining the maximum quadrature squeezing (min noise). Here, in the beam splitter approach, we minimize $\lambda_{\rm \scriptscriptstyle Simon}(\phi,{\rm sign}(c))$ with respect to the beam splitter angle $\phi$ and the ${\rm sign}(c)$ of the DGCZ criterion, in order to obtain the strongest single-mode nonclassicality condition. Now, one can observe that this minimization with respect to the beam splitter operator, actually, corresponds to finding the maximum quadrature-squeezing of a single-mode light by $x$-$p$ place rotations $\hat{a}_\phi=e^{i\phi}\hat{a}$. Such a minimization of $\lambda_{\rm \scriptscriptstyle Simon}(\phi,{\rm sign}(c))$ yields a simple  condition
\begin{equation}
\langle\hat{a}^\dagger\hat{a}\rangle \; < \;  |\langle\hat{a}^2\rangle|
\label{DGCZcond}
\end{equation}
for the presence of nonclassicality. For a Gaussian single-mode state Eq.~(\ref{DGCZcond}) is both a necessary and a sufficient condition for nonclassicality.

Repeating ourselves: what one performs in Eq.~(\ref{DGCZcond1}) is actually choosing the minimum noise direction among all possible quadrature-squeezing values in $(\Delta \hat{x}_\phi)^2$, see Sec. II.1 in Ref.~\cite{tasginAnatomy2019}. We kindly underline that the condition $\langle\hat{a}^\dagger\hat{a}\rangle <  |\langle\hat{a}^2\rangle|$ is obtained by performing optimization both on $\lambda_{\rm \scriptscriptstyle DGCZ}(c)$ criterion and with respect to the beam splitter parameters.

%%%%%%%%%%%%%%%%%%%%%%%%%%%%%%%%%%%%%%%%%%%%%%%%%%%%%%%%%%%%%%%%%%%%%%%%%%%%%%%%%%%%%%%%%%%%%%%%%%%%%%%%%%%
%%%%%%%%%%%%%%%%%%%%%%%%%%%%%%%%%%%%%%%%%%%%%%%%%%%%%%%%%%%%%%%%%%%%%%%%%%%%%%%%%%%%%%%%%%%%%%%%%%%%%%%%%%%
%%%%%%%%%%%%%%%%%%%%%%%%%%%%%%%%%%%%%%%%%%%%%%%%%%%%%%%%%%%%%%%%%%%%%%%%%%%%%%%%%%%%%%%%%%%%%%%%%%%%%%%%%%%
\section{Summary} \label{sec:summary}

In summary, we utilize the ``entanglement potential", introduced by Asboth {\it et al.}~\cite{EntanglementPot} for a single-mode nonclassical state, as a single-mode nonclassicality measure. Nonclassicality of a single-mode state can be quantified with the two-mode entanglement it generates at the beam splitter output. We carry out optimization over the parameters of both the beam splitter and the two-mode entanglement criteria.

We demonstrate that when the two-mode entanglement criterion of Duan-Giedke-Cirac-Zoller~\cite{Duan&Zoller2000} is employed at the beam splitter output, nonclassicality condition for the input single-mode state becomes the generalized quadrature-squeezing condition. This condition determines the optimum squeezing angle automatically. 

We obtain explicit matrix forms, in terms of $\langle\hat{a}^2\rangle$ and $\langle\hat{a}^\dagger \hat{a}\rangle$, of a nonclassicality measure for Gaussian single-mode states. We use the logarithmic negativity at the beam splitter output as the single-mode nonclassicality quantification. We measure the nonclassicality of quadrature-squeezed light and the nonclassicality generated in a second harmonic process in a damped nonlinear crystal. In the nonlinear crystal, we examine where the single mode nonclassicality becomes resistant (onsets) against the vacuum noises.

 We obtain a nonclassicality condition for general single-mode quantum states utilizing the two-mode entanglement criterion of Simon~\cite{SimonPRL2000}. We demonstrate that this condition not only successfully detects the single-mode nonclassicality of a superradiant light, but the violation of the condition also accompanies the strength of the superradiant phase.

%\bibliography{bibliography}

\begin{thebibliography}{0}%
\makeatletter
\providecommand \@ifxundefined [1]{%
 \@ifx{#1\undefined}
}%
\providecommand \@ifnum [1]{%
 \ifnum #1\expandafter \@firstoftwo
 \else \expandafter \@secondoftwo
 \fi
}%
\providecommand \@ifx [1]{%
 \ifx #1\expandafter \@firstoftwo
 \else \expandafter \@secondoftwo
 \fi
}%
\providecommand \natexlab [1]{#1}%
\providecommand \enquote  [1]{``#1''}%
\providecommand \bibnamefont  [1]{#1}%
\providecommand \bibfnamefont [1]{#1}%
\providecommand \citenamefont [1]{#1}%
\providecommand \href@noop [0]{\@secondoftwo}%
\providecommand \href [0]{\begingroup \@sanitize@url \@href}%
\providecommand \@href[1]{\@@startlink{#1}\@@href}%
\providecommand \@@href[1]{\endgroup#1\@@endlink}%
\providecommand \@sanitize@url [0]{\catcode `\\12\catcode `\$12\catcode
  `\&12\catcode `\#12\catcode `\^12\catcode `\_12\catcode `\%12\relax}%
\providecommand \@@startlink[1]{}%
\providecommand \@@endlink[0]{}%
\providecommand \url  [0]{\begingroup\@sanitize@url \@url }%
\providecommand \@url [1]{\endgroup\@href {#1}{\urlprefix }}%
\providecommand \urlprefix  [0]{URL }%
\providecommand \Eprint [0]{\href }%
\providecommand \doibase [0]{http://dx.doi.org/}%
\providecommand \selectlanguage [0]{\@gobble}%
\providecommand \bibinfo  [0]{\@secondoftwo}%
\providecommand \bibfield  [0]{\@secondoftwo}%
\providecommand \translation [1]{[#1]}%
\providecommand \BibitemOpen [0]{}%
\providecommand \bibitemStop [0]{}%
\providecommand \bibitemNoStop [0]{.\EOS\space}%
\providecommand \EOS [0]{\spacefactor3000\relax}%
\providecommand \BibitemShut  [1]{\csname bibitem#1\endcsname}%
\let\auto@bib@innerbib\@empty
%</preamble>
\end{thebibliography}%


\begin{thebibliography}{99}


%BIBLIOGRAPHY---BIBLIOGRAPHY---BIBLIOGRAPHY---BIBLIOGRAPHY---BIBLIOGRAPHY---BIBLIOGRAPHY---BIBLIOGRAPHY---
%%%%%%%%%%%%%%%%%%%%%%%%%%%%%%%%%%%%%%%%%%%%%%%%%%%%%%%%%%%%%%%%%%%%%%%%%%%%%%%%%%%%%%%%%%%%%%%%%%%%%%%%%%%%



%Entanglement by a beam splitter: Nonclassicality as a prerequisite for entanglement
\bibitem{Kim&Knight2002} M. S. Kim, W. Son, V. Buzek, and P. L. Knight, Phys. Rev. A {\bf 65}, 032323 (2002).

%Theorem for the beam splitter entangler
\bibitem{BSentanglement} W. Xiang-bin, Phys. Rev. A {\bf 66}, 024303 (2002).

%A quantum characterization of classical radiation
\bibitem{aharanov1966} Y. Aharonov, D. Falkoff, E. Lerner, and H. Pendleton, Ann. Phys. {\bf 39}, 498 (1966).


%Teleporting an unknown quantum state via dual classical and Einstein-Podolsky-Rosen channels
\bibitem{teleport1} C. H. Bennett, G. Brassard, C. Crepeau, R. Jozsa, A. Peres, and W. K. Wootters, Phys. rev. Lett. {\bf 70}, 1895 (1993).

%Teleportation of Continuous Quantum Variables
\bibitem{teleport2} S. L. Braunstein and H. J. Kimble, Phys. Rev. Lett. {\bf 80}, 869 (1998).

%Long-distance quantum communication with atomic ensembles and linear optics
\bibitem{teleport3} L.-M. Duan, M. D. Lukin, J. I. Cirac, and P. Zoller, Nature {\bf 414}, 413 (2001).


%Measuring nanomechanical motion with an imprecision below the standard quantum limit
\bibitem{measurementSQL1} G. Anetsberger, E. Gavartin, O. Arcizet, Q. P. Unterreithmeier, E. M. Weig, M. L. Gorodetsky, J. P. Kotthaus, and T. J. Kippenberg, Phys. Rev. A {\bf 82}, 061804(R) (2010).

%Nanomechanical motion measured with an imprecision below that at the standard quantum limit
\bibitem{measurementSQL2} J. D. Teufel, T. Donner, M. A. Castellanos-Beltran1, J. W. Harlow1, and K. W. Lehnert, Nature Nanotechnology {\bf 4}, 820 (2009).

%Quantum-measurement backaction from a Bose-Einstein condensate coupled to a mechanical oscillator
\bibitem{measurementSQL3} S. K. Steinke, S. Singh, M. E. Tasgin, P. Meystre, K. C. Schwab, and M. Vengalattore, Phys. Rev. A {\bf 84}, 023841 (2011). 

%Quantum-mechanical limitations in macroscopic experiments and modern experimental technique
\bibitem{measurementSQL4} V. B Braginskii and Y. I. Vorontsov,  Sov. Phys. Usp. {\bf 17}, 644 (1975).

\bibitem{LIGO2013} Aasi, Junaid, B. P. Abbott, Richard Abbott, Thomas Abbott, M. R. Abernathy, Kendall Ackley, Carl Adams et al. "Advanced LIGO. {\it Classical and quantum gravity} {\bf 32}, 074001 (2015).



%Negativity of theWigner function as an indicator of non-classicality
\bibitem{Wnegativity1} A. Kenfack and K. Zyczkowski, J. Opt. B {\bf 396}, 396 (2004).
%Continuous-variable teleportation of a negative Wigner function
\bibitem{Wnegativity2} L. Mista Jr., R. Filip, and A. Furusawa, Phys. Rev. A {\bf 82}, 012322 (2010).

%Probing the Negative Wigner Function of a Pulsed Single Photon Point by Point
\bibitem{WignerfnxExperiment} K. Laiho, K. N. Cassemiro, D. Gross, and C. Silberhorn, Phys. Rev. Lett. {\bf 105}, 253603 (2010).

%\bibitem{measurementSQL3} S.K. Steinke, S. Singh, M.E. Tasgin, P. Meystre, K.C. Schwab, M. Vengalattore, Phys. Rev. A {\bf 84}, 023841 (2011).


\bibitem{Pnegativity} A. Ferraro and M. G. A. Paris, Phys. Rev. Lett. {\bf 108}, 260403 (2012).
%Equivalence of Semiclassical and Quantum Mechanical Descriptions of Statistical Light Beams
\bibitem{SudarshanP} E. C. G. Sudarshan, Phys. Rev. Lett. {\bf 10}, 277 (1963).



\bibitem{tasgin2017many} Mehmet Emre Tasgin, {\it Many-particle entanglement criterion for superradiant-like states}, Phys. Rev. Lett. {\bf 119}, 033601 (2017).

\bibitem{tasgin2015HP} M. E. Tasgin, {\it Single-mode nonclassicality criteria via Holstein-Primakoff transformation}, arXiv:1502.00988.

\bibitem{Klauder1985} J. R. Klauder and B. Skagerstam, {\it Applications in physics and mathematical physics}, World Scientific, Singapore (1985).
\bibitem{Radcliffe1971} J. M. Radcliffe, {\it Some properties of coherent spin states}, Journal of Physics A: General Physics {\bf 4}, 313 (1971).



%Peres-Horodecki Separability Criterion for Continuous Variable Systems
\bibitem{SimonPRL2000} R. Simon, Phys. Rev. Lett. {\bf 84}, 2726 (2000).

%Extremal entanglement and mixedness in continuous variable systems
\bibitem{AdessoPRA2004} G. Adesso, A. Serafini, and F. Illuminati, Phys. Rev. A {\bf 70}, 022318 (2004).

%Computable measure of entanglement
\bibitem{WernerPRA2002} G. Vidal and R. F. Werner, Phys. Rev. A {\bf 65}, 032314 (2002).

%Inseparability Criterion for Continuous Variable Systems
\bibitem{Duan&Zoller2000} Lu-Ming Duan, G. Giedke, J. I. Cirac, and P. Zoller, Phys. Rev. Lett. {\bf 84}, 2722 (2000).


%Computable measure of Nonclassicality for Light
\bibitem{EntanglementPot} J. K. Asboth, J. Calsamiglia, and H. Ritsch, Phys. Rev. Lett. {\bf 94}, 173602 (2005).



%Entanglement and nonclassicality for multimode radiation-field states
\bibitem{SimonPRA2011} J. S. Ivan, S. Chaturvedi, E. Ercolessi, G. Marmo, G. Morandi, N. Mukunda, R. Simon, Phys. Rev. A {\bf 83}, 032118 (2011).


%Unified quantification of nonclassicality and entanglement
\bibitem{Vogel&Sperling2014} W. Vogel and J. Sperling, Phys. Rev. A {\bf 89}, 052302 (2014).

%Witnessing the degree of nonclassicality of light
\bibitem{Mraz&Hage2014} M. Mraz, J. Sperling, W. Vogel, and B. Hage, Phys. Rev. A {\bf 90}, 033812 (2014).



\bibitem{HZPRA2006} Mark Hillery and M. Suhail Zubairy, {\it Entanglement conditions for two-mode states: Applications} Phys. Rev. A {\bf 74}, 032333 (2006).


\bibitem{Plenio2005} Martin B Plenio, {\it Logarithmic negativity: A full entanglement monotone that is not convex}, Phys. Rev. Lett. {\bf 95}, 090503 (2005).


\bibitem{MGAParisPRA2015} M. Brunelli, C. Benedetti, S. Olivares, A. Ferraro, and Matteo GA Paris. {\it Single-and two-mode quantumness at a beam splitter.} Phys. Rev. A {\bf 91}, 062315 (2015).

\bibitem{NhaSciRep2019} J. Park, J. Lee, and Hyunchul Nha. {\it Entropic nonclassicality and quantum non-Gaussianity tests via beam splitting.}, Scientific Reports {\bf 9}, 1-13 (2019).



\bibitem{simonPRA1994} R. Simon, N. Mukunda, and Biswadeb Dutta. {\it Quantum-noise matrix for multimode systems: U (n) invariance, squeezing, and normal forms.}, Phys. Rev. A {\bf 49} , 1567 (1994).










%%Single-mode nonclassicality criteria via Holstein-Primakoff transformation
%\bibitem{metasginHPtransformation} M. E. Tasgin, "{\it Single-mode nonclassicality criteria via Holstein-Primakoff transformation}" (2014).

%%Optical Coherence book
%\bibitem{mandelwolf} L. Mandel and E. Wolf, {\it Optical Coherence and Quantum Optics}, (Cambridge University Press, Cambridge, 1995).


%Chaos and the quantum phase transition in the Dicke model
\bibitem{Emary&BrandesPRE2003} C. Emary and T. Brandes, Phys. Rev. E {\bf 67}, 066203 (2003).

\bibitem{BrandesPRA2005Q} N. Lambert, C. Emary, and T. Brandes. "Entanglement and entropy in a spin-boson quantum phase transition." Physical Review A 71, no. 5 (2005): 053804.

%%Holstein&Primakoff Transformation
%\bibitem{Holstein&Primakoff} T. Holstein and H. Primakoff, Phys. Rev. {\bf 58}, 1098 (1949)
%
%%%Coherent States: Applications in Physics and Mathematical Physics
%\bibitem{CoherentStatesbook}  J. R. Klauder and B. S. Skagerstam, {\it Coherent States: Applications in Physics and Mathematical Physics}, (World Scientific Pub Co. Inc., 1985).
%
%%Some properties of coherent spin states
%\bibitem{Radcliffe1971} J. M. Radcliffe, J. Phys. A: Gen. Phys. {\bf 4}, 313 (1971). 
%



%Quantum Optics
\bibitem{Scullybook} M. O. Scully and M. S. Zubairy,  {\it Quantum Optics}, (Cambridge University Press, Cambridge, 1997).
%
%Optomechanical Entanglement between a Movable Mirror and a Cavity Field
\bibitem{optomechEnt1} D. Vitali, S. Gigan, A. Ferreira, H. R. B\"{o}hm, P. Tombesi, A. Guerreiro, V. Vedral, A. Zeilinger, and M. Aspelmeyer, Phys. Rev. Lett. {\bf 98}, 030405 (2007).
%
%%Robust entanglement of a micromechanical resonator with output optical fields
%\bibitem{optomechEnt2} C. Genes, A. Mari, P. Tombesi, and D. Vitali, Phys. Rev. A {\bf 78}, 032316 (2008).

%%Quantum measurement backaction from a BEC coupled to a mechanical oscillator,
%\bibitem{MeystreBackaction} S.K. Steinke, S. Singh, M.E. Tasgin, P. Meystre, K.C. Schwab and M. Vengalattore, Phys. Rev. A {\bf 84}, 023841 (2011).  

%%Mutual emergence of noncausal optical response and nonclassicality in a optomechanical system
%\bibitem{metasginnoncausal} D. Tarhan, O. E. Mustecaplioglu and M. E. Tasgin, "{\it Mutual emergence of noncausal optical response and nonclassicality in a optomechanical system}" (2014).


%%Entanglement Conditions for Two-Mode States
%\bibitem{Hillery&Zubairy2006} M. Hillery and M. S. Zubairy, Phys. Rev. Lett. {\bf 96}, 050503 (2006).
%

%Quantum correlated light pulses from sequential superradiance of a condensate
\bibitem{metasginEndfire} M.E. Tasgin, M.O. Oktel, L. You, and O.E. Mustecaplioglu, Phys. Rev. A {\bf 79}, 053603 (2009).  

%%Entanglement in a parametric converter
%\bibitem{parametricPhase1} S.-Y. Lee1, S. Qamar, H.-W. Lee, and M Suhail Zubairy, J. Phys. B: At. Mol. Opt. Phys. {\bf 41}, 145504 (2008).

%%Experimental investigation about the influence of pump phase noise on phase-correlation of output optical fields from a non-degenerate parametric oscillator
%\bibitem{parametricPhase2} D. Wang, Y. Shang, Z. Yan, W. Wang, X. Jia, C. Xie and K. Peng, Europhys. Lett. {\bf 82}, 24003 (2008).
%
%%Influence of pump-phase fluctuations on entanglement generation using a correlated spontaneous-emission laser
%\bibitem{parametricPhase3} S. Qamar, H. Xiong, and M. S. Zubairy, Phys. Rev. A {\bf 75}, 062305 (2007).

%Superradiant Rayleigh scattering from a Bose-Einstein condensate
\bibitem{KetterleSience1999} S. Inouye, A.P. Chikkatur, D.M. Stamper-Kurn, J. Stenger, D.E. Pritchard, and W. Ketterle, Science {\bf 285}, 571 (1999).

%%
\bibitem{Dicke1954} R. H. Dicke, Phys. Rev. {\bf 93}, 99 (1954).

\bibitem{Skribanowitz1973} N. Skribanowitz, I. P. Herman, J. C. MacGillivray, and M. S. Feld, Phys. Rev. Lett. {\bf 30}, 309 (1973).

%Entanglement and the Phase Transition in Single-Mode Superradiance
\bibitem{BrandesPRL2004} N. Lambert, C. Emary, and T. Brandes, Phys. Rev. Lett. {\bf 92}, 073602 (2004).


%Pairwise entanglement in symmetric multi-qubit systems
\bibitem{symmetricDicke} X. Wang and K. M{\o}lmer, Eur. Phys. J. D {\bf 18}, 385 (2002).

\bibitem{tasginAnatomy2019} M. E. Tasgin, {\it Anatomy of entanglement and nonclassicality criteria},	arXiv:1901.04045 (2019). 



%
\bibitem{Vitali2008PRA} G. Claudiu, A. Mari, P. Tombesi, and D. Vitali. {\it Robust entanglement of a micromechanical resonator with output optical fields.}, Phys. Rev. A {\bf 78}, 032316 (2008).


\bibitem{VitaliPRL2007} D. Vitali, S. Gigan, A. Ferreira, H. R. Bohm, P. Tombesi, A. Guerreiro, V. Vedral, A. Zeilinger, and M. Aspelmeyer. {\it Optomechanical entanglement between a movable mirror and a cavity field.}, Phys. Rev. Lett. {\bf 98}, 030405 (2007).

\bibitem{QuantumNoiseBook} C. Gardiner and P. Zoller, {\it Quantum noise: a handbook of Markovian and non-Markovian quantum stochastic methods with applications to quantum optics.} Springer Science \& Business Media, 2004.




%
%
%
%%Holographic Dual of an Einstein-Podolsky-Rosen Pair has a Wormhole
%\bibitem{entanWorm1} K. Jensen and A. Karch, PRL 111, 211602 (2013).
%
%%Holographic Schwinger Effect and the Geometry of Entanglement
%\bibitem{entanWorm2} J. Sonner, Phys. Rev. Lett. {\bf 111}, 211603 (2013).
%
%%Cool horizons for entangled black holes
%\bibitem{entanWorm3} J. Maldacena and L. Susskind, arXiv:1306.0533v2 (2013).
%
%\bibitem{PS1} We note that the superluminal communication, we mention here, is not be confused with superluminal group velocities \cite{superluminalexp} which are shown to be unreliable \cite{superluminal1,superluminal2,superluminal3}.
%
%\bibitem{superluminalexp}  L. J.Wang, A. Kuzmich, and A. Dogariu, Nature (London) {\bf 406}, 277 (2000).
%
%%Direct Observation of a Pulse Peak Using a Peak-Removed Gaussian Optical Pulse in a Superluminal Medium
%\bibitem{superluminal1} M. Tomita, H. Amano, S. Masegi, and A. I. Talukder, Phys. Rev. Lett. {\bf 112}, 093903 (2014).
%
%%Testing the reliability of a velocity definition in a dispersive medium
%\bibitem{superluminal2} M. E. Tasgin, Phys. Rev. A {\bf 86}, 033833 (2012).
%
%%On the supraluminal group velocity and the transmission of information
%\bibitem{superluminal3} S. N. Molotkov, JETP Lett. {\bf 91}, 693 (2010).
%
%
%\bibitem{JacksonEMT} J. D. Jackson, {\it Classical Electrodynamics}, 3rd ed. (Wiley, New York, 1998).
%

%%Superluminal Behavior, Causality and Instability
%\bibitem{SusskindSuperluminal} Y. Aharonov, A. Komar, and L. Susskind, Phys. Rev. {\bf 182}, 1400 (1969).
%
%%Quantum field theory cannot provide faster than light communication
%\bibitem{nosuperluminal_communication} P. H. Eberhard, R. R. Ross, Found. Phys. Lett. {\bf 2}, 127 (1989).
%
%
%%Gaussian pure states in quantum mechanics and the symplectic group
\bibitem{Simon1988} R. Simon, E. C. G. Sudarshan, and N. Mukunda, Phys. Rev. A {\bf 37}, 3028 (1988).
%
%
%
%%'Superfluid spacetime' points to unification of physics
%\bibitem{NatureNews} C. Moskowitz, {\it 'Superfluid spacetime' points to unification of physics}, Nature News, 19 June 2014.
%
%%Astrophysical Constraints on Planck Scale Dissipative Phenomena
%\bibitem{MaccionePRL2014} S. Liberati and L. Maccione, Phys. Rev. Lett. {\bf 112}, 151301 (2014). 
%










%%
%
%
%
%

%


\end{thebibliography}

%\newpage

%

%%%%%%%%%%%%%%%%%%%%%%%%%%%%%%%%%%%%%%%%%%%%%%%%%%%%%%%%%%%%%%%%%%%
\begin{acknowledgements}
I thank Nathan Killoran for his leading comments.  I thank \"{O}zg\"{u}r E. M\"{u}stecapl{\i}o\u{g}lu and G\"{u}rsoy B. Akg\"{u}\c{c} for illuminating and leading discussions. I acknowledge help from Shailendra K. Singh on verifying my derivations. We acknowledge support  from T\"{U}B\.{I}TAK-KAR\.{I}YER  Grant No.  112T927,  T\"{U}B\.{I}TAK-1001  Grant No.  114F170 and 117F118, and Hacettepe University BAP-6091 Grant No. 014G602002. 
\end{acknowledgements}

%%%%%%%%%%%%%%%%%%%%%%%%%%%%%%%%%%%%%%%%%%%%%%%%%%%%%%%%%%%%%%%%%%%

%%%%%%%%%%%%%%%%%%%%%%%%%%%%%%%%%%%%%%%%%%%%%%%%%%%%%%%%%%%%%%%%%%
%%%%%%%%%%%%%%%%%%%%%%%%%%%%%%%%%%%%%%%%%%%%%%%%%%%%%%%%%%%%%%%%%%
%%%%%%%%%%%%%%%%%%%%%%%%%%%%%%%%%%%%%%%%%%%%%%%%%%%%%%%%%%%%%%%%%%%
%\begin{appendix}
%\section{Equations of Motion}
%
%
%
%\end{appendix}
%%%%%%%%%%%%%%%%%%%%%%%%%%%%%%%%%%%%%%%%%%%%%%%%%%%%%%%%%%%%%%%%%%
%%%%%%%%%%%%%%%%%%%%%%%%%%%%%%%%%%%%%%%%%%%%%%%%%%%%%%%%%%%%%%%%%%
\end{document}